# MIMICS-Duo: Offline & Online Evaluation of Search Clarification

Leila Tavakoli
RMIT University
Australia
leila.tavakoli@rmit.edu.au

Johanne R. Trippas
University of Melbourne
Australia
johanne.trippas@unimelb.edu.au

Hamed Zamani
University of Massachusetts Amherst
United States
zamani@cs.umass.edu

Falk Scholer
RMIT University
Australia
falk.scholer@rmit.edu.au

Mark Sanderson
RMIT University
Australia
mark.sanderson@rmit.edu.au

## ABSTRACT

Asking clarification questions is an active area of research; however, resources for training and evaluating search clarification methods are not sufficient. To address this issue, we describe MIMICS-Duo, a new freely available dataset of 306 search queries with multiple clarifications (a total of 1,034 query-clarification pairs). MIMICS-Duo contains fine-grained annotations on clarification questions and their candidate answers and enhances the existing MIMICS datasets by enabling multi-dimensional evaluation of search clarification methods, including online and offline evaluation. We conduct extensive analysis to demonstrate the relationship between offline and online search clarification datasets and outline several research directions enabled by MIMICS-Duo. We believe that this resource will help researchers better understand clarification in search.

## CCS CONCEPTS

• **Information systems** → **Search interfaces**.

## KEYWORDS

Search clarification dataset, Online/Offline evaluation, Clarification selection

## 1 INTRODUCTION

Retrieving documents for ambiguous, faceted, incomplete, or complex queries can be improved by clarifying user information needs. Recent studies showed that clarification in search could introduce functional and emotional benefits for users [38] and the most common mixed-initiative interaction type in conversational systems is clarification [4, 41]. Advancing state of the art in generating, selecting, and presenting clarification questions is tightly coupled with developing effective evaluation methodologies and resources for their quantitative assessment. Examining existing public resources for search clarification demonstrated that they are not sufficient for a multi-dimensional evaluation of search clarification methods. This paper bridges this gap by introducing a new dataset, MIMICS-Duo,[1] that enables both online and offline evaluation of methods for clarification selection and generation. The queries in MIMICS-Duo are sampled from MIMICS-ClickExplore [39], a large-scale dataset for search clarification consisting of online signals, such as user engagement based on click-through rate (CTR). By conducting multi-dimensional manual annotation for over one thousand query-clarification pairs (i.e. overall quality labelling for panes and individual candidate answers, offline rating of panes and aspect labelling, MIMICS-Duo together with MIMICS-ClickExplore enables both online and offline evaluation of search clarification for over 300 search queries.

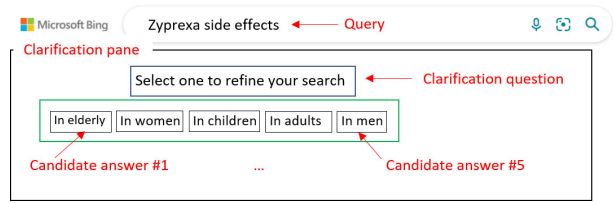

**Figure 1: A query and a clarification pane.**

MIMICS-Duo can be used for training and evaluating many search clarification tasks: generating clarification questions; ranking clarification panes (Figure 1); re-ranking candidate answers; unbiased click models and user engagement prediction for clarification; and analyzing user interaction with search clarification.

This paper details existing resources and their limitations (Section 3) and the formation of MIMICS-Duo (Section 4). We analyze the properties of MIMICS-Duo (Section 5), which indicates that no relationship exists between CTR level as an indicator of clarification question engagement (online evaluation) and manual labels (offline evaluation). The MIMICS-Duo dataset helps us establish the relationships between different aspects of clarification panes which can be used for further improvement of generating and asking clarification models. Lastly, we discuss potential future research pathways (Section 6) and finish with conclusions (Section 7).

---
[1]MIMICS-Duo is available at https://github.com/Leila-Ta/MIMICS-Duo

## 2 RELATED WORK AND RESOURCES

**Clarification in Information Retrieval.** Clarification questions are examined in several fields: dialogue systems [2, 10, 16], community question answering (CQA) [6, 18, 28, 32], conversational search systems [3, 10, 38, 42], and speech recognition [31]. Studies of clarification in conversational search can be divided into clarification question *generation* and *selection*.

For clarification question generation, Cao et al. [9] proposed a model which could produce questions with various levels of specificity. Zamani et al. [38] proposed supervised and reinforcement learning models for generating clarification questions in a search engine. Dhole [10] used an existing question generator and a sentence similarity model to generate discriminative questions to resolve intent ambiguity in dialogue systems. A challenge of generation is the possibility of different user intents, which makes generating one generic clarification question per context unsuccessful. To resolve this challenge, Zhang and Zhu [42] proposed a model that predicted keywords focusing on the specific aspects of the question and produced multiple keyword groups for generation diversity.

For clarification question selection, Rao [26] and Rao and Daumé III [27] built a neural network model for asking clarification questions. Asking clarification questions in open-domain information-seeking conversational systems was the focus of the study conducted by Aliannejadi et al. [3]. They proposed a neural question selection model capable of asking clarification questions that can address users' information needs. Zamani et al. [40] focused on learning representations for clarification questions from user interaction. The model showed successful performance on re-ranking automatically generated clarification questions for a given query.

For research datasets, we can divide resources into two main categories: CQA clarification datasets and search clarification datasets. Among the clarification datasets in CQA, Rao [26] and Rao and Daumé III [27] extracted clarifications from three domains of *askubuntu*, *unix* and *superuser*. Braslavski et al. [6] used two Stack-Exchange sites of *Home Improvements* (DIY) and *Arqade* (GAMES) to build a clarification dataset. Xu et al. [36], Kumar et al. [17] and Tavakoli et al. [32] were other research groups that created their clarification datasets using CQA and KBQA (Knowledge-Based Question Answering). Some other researchers such as Rao and Daumé III [28], Zhang and Zhu [42], and Majumder et al. [21] investigated generating clarification question models on the Amazon Review dataset [23, 24]. However, CQA datasets are of limited use in the search clarification domain. The datasets record human interactions, while in a conversational search system, a human interacts with a machine. The nature of the query and the information need is also different in search clarification compared to a community forum (i.e., synchronous vs. asynchronous).

Aliannejadi et al. [3] collected a clarification question dataset through crowdsourcing named Qulac. It contains 200 queries from the TREC Web Track and human-generated clarification questions. Inspired by Qulac, Aliannejadi et al. [1, 2] crowdsourced new datasets to study clarification questions that were suitable for conversational settings and in open domain dialogues focusing on single and multi-turn conversations.

We focus on MIMICS, the largest search clarification dataset extracted from web search query logs. MIMICS includes two subsets of user interactions with search clarification in a commercial search engine and a subset containing quality labels for clarification panes collected through manual annotation. Compared to other datasets, MIMICS contains realistic queries, is comprehensive and covers a wide range of clarification types and user interaction signals. Table 1 provides a comparison between available datasets and MIMICS-Duo which will be discussed further in Section 4.

**Online and Offline Evaluations.** Offline evaluations provide a low-cost methodology to predict the performance of models and insight into whether it is worth testing on the more expensive online evaluation. However, literature reviews show that there are substantial discrepancies between the offline and online performance of models [11, 14, 29, 43]. For instance, Beel et al. [5] found that results of offline and online evaluations of recommenders often contradict each other as offline evaluations normally ignore human factors. This was also highlighted by Yi et al. [37] that stated offline metrics can be misleading. In another study, Garcin et al. [14] investigated news recommenders and showed that in an offline setting, recommending popular stories is a winning strategy, but online, it was the poorest. On the other side, Zheng et al. [43] and later Garcin et al. [14] concluded that the click-through rate (CTR), an adopted and widely accepted metric in online evaluations, overestimates the impact of popular items. In fact, recommending items with higher CTR does not necessarily imply higher relevance of two items, and factors like item popularity, item serendipity or the placement/order of recommendations may also influence a user's click behaviour. Apart from potential factors mentioned here, Liu et al. [19] stated that the definition of satisfaction is rather subjective and different users may have different opinions in satisfaction judgement. Similarly, Mao et al. [22] looked at this problem from another angle and showed that relevance, as annotated by external assessors, may not necessarily mean usefulness and satisfaction appreciated by users. They showed that a measure based on usefulness had a better correlation with user satisfaction than relevance.

The literature review shows that although generating and asking clarification questions in search engines have advanced noticeably over the last few years, our lack of knowledge about online and offline evaluations in search clarification calls into question the models' performance, which makes the application of search clarification limited. Available clarification questions datasets are either created based on the user interaction signals such as click-through rate or collected through manual annotation. Therefore, a clear relationship between online and offline evaluations cannot be established. This is the missing link that the MIMICS dataset, as the largest and the most realistic search clarification data collection, cannot yet address, although inspired by several other studies [15, 20, 30, 34]. We aim to resolve this shortcoming in search clarification by introducing MIMICS-Duo, a balanced dataset that benefits from user interaction signals while providing insightful information about the characteristics of clarification questions.

## 3 MOTIVATION

Methods for generating and selecting conversational search clarification questions [3, 38, 40, 41], have been assessed using either online or offline evaluations, while the differences between the two evaluation methods have been overlooked. Progress in this important research topic relies on thorough experimentation by considering both online and offline evaluation methodologies. We conduct a series of preliminary experiments on the existing MIMICS dataset [39], to illustrate the limitations of MIMICS, such as a very low overlap between its online and offline components, that prevent researchers from performing comprehensive analysis.

Table 1: Statistics of the clarification datasets.

| Dataset Type | CQA[1] | Search Clarification | | |
|---|---|---|---|---|
| Dataset Name | ClarQ, ... | Qulac | MIMICS | MIMICS-Duo |
| Source | StackExchange, Amazon, ... | TREC Web Track | Search Engine | Search Engine |
| # Queries | various (37K - 2M) | 198 | 450K | 306 |
| Clarification Type | Post and comment | Human generated | Machine Generated | Machine Generated |
| Evaluation Method | Offline | Offline | Online or Offline | Online & Offline |
| Overlap between online & offline | Null | Null | 106 query-clarification pair | 1,034 query-clarification pair |

[1] Braslavski et al. [6], Rao [26], Rao and Daumé III [27], Tavakoli et al. [32], Xu et al. [36]

### 3.1 Clarification Selection

As a case study for highlighting the differences between online and offline evaluation, we focus on the task of clarification re-ranking or selection [3, 18, 25, 27, 40], i.e. aiming to improve the quality of which clarification from among a set of options is presented. Using a wide range of ranking models, we demonstrate their behaviour in offline versus online evaluation setups.

*3.1.1 Data Collection and Pre-Processing.* We use the MIMICS dataset [39], which consists of three subsets:

**MIMICS-Click:** Includes over 400,000 unique queries, their associated clarification panes, and the corresponding aggregated user interaction signals. Each data point in MIMICS-Click includes a query-clarification pair, an impression level (low, medium, or high), an engagement level (i.e., an integer between 0 to 10 presenting the level of total engagement received by users in terms of click-through rate), and the conditional click probability for each individual candidate answer.

**MIMICS-ClickExplore:** Includes over 60,000 unique queries. Each query is associated with multiple clarification panes in addition to the user interaction signals, similar to MIMICS-Click.

**MIMICS-Manual:** Includes over 2,000 unique search queries with multiple clarification panes, landing results pages and manually annotated quality labels (2 Good, 1 Fair, or 0 Bad) based on fluency, grammar, and clarification accuracy.

We focus on MIMICS-ClickExplore and MIMICS-Manual as they include multiple clarification panes per query. MIMICS-ClickExplore can be seen as a dataset for online evaluation because it contains signals based on user engagement with *Bing* clarification panes. However, MIMICS-Manual contains manual expert annotations for clarification quality, and it follows the traditional approach for offline evaluation.

In MIMICS-ClickExplore, there are a few identical query-clarification pairs with different impression or engagement levels. In these cases, we randomly retain one clarification pane and discard the rest. As a result of this process, 708 queries are left with only one clarification pane that cannot be used for the re-ranking task. Therefore, we skip these queries in our experiments. We also remove queries with the clarification panes that received the same engagement level since re-ranking these clarification panes has no difference. This pre-processing leaves 61,222 unique queries from MIMICS-ClickExplore.

For MIMICS-Manual, we consider the overall quality label for each clarification pane. The majority of queries in this dataset have only one clarification pane, or all their clarification panes receive the same quality score. Consequently, this dataset only contains 66 queries useful for the clarification re-ranking (selection) task, a major limitation of MIMICS-Manual.

*3.1.2 Task Formulation and Experimental Setup.* To investigate the relationship between online and offline evaluation in search clarification, we conduct experiments using learning to rank (LTR) models for re-ranking clarification panes in response to each query, including *MART* [13], *RankNet* [8], *RankBoost* [12], *Coordinate Ascent* [33], *LambdaMart* [35], and *RandomForests* [7]. We use five-fold cross-validation in our experiments. An extensive set of features and their combinations are explored – 110 features in total, grouped into five categories as shown in Table 2. The features are linearly normalized based on their min/max values. The code for extracting these features and their descriptions are available on GitHub[2].

Since users typically are only shown one clarification pane; thus, we use P@1 and mean reciprocal rank (MRR) for evaluation. MRR is calculated by the position of the top-rated document, here clarification pane, according to the engagement level or quality, depending on the dataset being used for evaluation. Clarification pane labels are decided based on engagement levels (on MIMICS-ClickExplore) or overall option quality (on MIMICS-Manual).

*3.1.3 Experimental Results.* In our first set of experiments, we train LTR models on MIMICS-ClickExplore and evaluate them on both MIMICS-ClickExplore and MIMICS-Manual. According to the results shown in Table 3, the performance of the different models on the MIMICS-Manual dataset varies substantially across models (in terms of both P@1 and MRR), while this is not the case for MIMICS-ClickExplore. On MIMICS-ClickExplore, RankBoost and Coordinate Ascent perform similarly, while there is a substantial difference between their performance on MIMICS-Manual. On the other hand, RandomForests performs better than Coordinate Ascent on MIMICS-Manual, which is not the case on MIMICS-ClickExplore. We carried out a t-test between the model effectiveness scores on the ClickExplore and Manual collections, respectively, with a threshold of $p < 0.05$ to determine significance. For ClickExplore,

[2] https://github.com/Leila-Ta/Clarification-LTR-Features

Table 2: LTR features and feeding inputs.

| Feature | Input 1 | Input 2 | # of Features |
|---|---|---|---|
| - TF-IDF using Cosine Similarity<br>- BM25 Similarity<br>- Overall Term-Matching | Query<br>Document Title<br>Related search<br>Video title<br>Video description<br>Snippet | Clarification pane | 18 |
| | | Clarification question<br>+<br>Candidate answer # [1, ..., 5] | 18@5: 90 |
| Number of candidate answers for each clarification question | Not applicable | Not applicable | 1 |
| Is a clarification question a question or a statement? | Question: 1<br>Statement: 0 | Not applicable | 1 |

11 of 15 pairwise tests show significant differences for P@1; and 8 of 15 pairwise tests show significant differences for MRR. Similar issues arise when the models are trained on MIMICS-Manual and then evaluated on the MIMICS-ClickExplore and MIMICS-Manual datasets, as shown in Table 4. For example, RankNet performs relatively well on MIMICS-Manual, while it shows the poorest performance on MIMICS-ClickExplore. The number of significant pairwise differences also shows variations, with 2 of 6 for P@1 on ClickExplore and Manual and two versus 1 for MRR on ClickExplore and Manual, respectively, when the training dataset changed for a model. Overall, both the specific system performance rankings and the sensitivity of the two collections differ.

Differences between using the online and offline datasets are further highlighted when examining the weight of each feature learned by an LTR model from each dataset. We focus on RankBoost, which has produced the best overall performance in several cases. Similar observations hold for the top features selected by other models as well. Table 5 shows the top 10 features with the highest weights according to RankBoost when it was trained on the MIMICS-ClickExplore and MIMICS-Manual datasets, respectively. It is striking that only two out of ten features are shared across the datasets, suggesting that RankBoost learned substantially different ranking functions from MIMICS-ClickExplore and MIMICS-Manual, even though the available features were the same.

Table 3: Performance of LTR models trained on MIMICS-ClickExplore for clarification selection (re-ranking), significance test results are explained in the text.

| Model | P@1 | | MRR | |
|---|---|---|---|---|
| Testing Dataset | ClickExp. | Manual | ClickExp. | Manual |
| MART | 0.415 | 0.394 | 0.667 | 0.697 |
| RankNet | 0.417 | 0.409 | 0.668 | 0.705 |
| RankBoost | 0.444 | 0.606 | 0.683 | 0.803 |
| Coordinate Ascent | 0.444 | 0.424 | 0.683 | 0.712 |
| LambdaMART | 0.424 | 0.561 | 0.671 | 0.780 |
| RandomForests | 0.417 | 0.455 | 0.667 | 0.727 |

## 3.2 Limitations of Existing Resources

Through the task of clarification selection, we demonstrated that existing resources are not sufficient for a full exploration of online versus offline evaluation in search clarification. To the best of our knowledge, MIMICS is the only data collection that provides both online and offline signals for evaluating search clarification and has enabled substantial advances in this space; however, its limitations are a barrier to advancing other investigations into search clarification. In particular, the nature of these datasets might have contributed to the different behaviours observed in the reported experiments. MIMICS-ClickExplore contains over 60K queries, while MIMICS-Manual contains only 66 unique queries, usable for comparison. The number of available clarification panes per query is also very different in the two datasets, and more importantly, the diversity of the quality labels provided in MIMICS-Manual is low. However, these are not just the only limitations. A close look at both datasets shows that there are only 106 query-clarification pairs shared between MIMICS-Manual and MIMICS-ClickExplore. This includes tied query-clarification pairs that have the same engagement level, and we removed them from our experiment. Therefore, drawing robust conclusions about the impact of online and offline evaluations in search clarification is not possible using available datasets as there are not many query-clarification pairs that have both online and offline information. Therefore, developing a dataset that enables researchers to perform thorough online and offline evaluations is highly important and motivated us to create the MIMICS-Duo dataset.

Table 4: Performance of LTR models trained on MIMICS-Manual for clarification selection (re-ranking), significance test results are explained in the text.

| Model | P@1 | | MRR | |
|---|---|---|---|---|
| Testing Dataset | ClickExp. | Manual | ClickExp. | Manual |
| MART | 0.425 | 0.485 | 0.673 | 0.742 |
| RankNet | 0.404 | 0.530 | 0.660 | 0.765 |
| RankBoost | 0.425 | 0.439 | 0.672 | 0.720 |
| Coordinate Ascent | 0.420 | 0.424 | 0.669 | 0.712 |
| LambdaMART | 0.411 | 0.439 | 0.664 | 0.720 |
| RandomForests | 0.417 | 0.561 | 0.667 | 0.780 |

## 4 METHODOLOGY

To create MIMICS-Duo that overcomes the shortcoming of the current search clarification datasets, we conducted online experiments[3]

---
[3] Reviewed and approved according to Anonymous University IRB procedures for research involving human subjects.

Table 5: The top 10 features with the highest weight learned by RankBoost from MIMICS-ClickExplore and MIMICS-Manual.

| Training on MIMICS-ClickExplore (online) | Training on MIMICS-Manual (offline) |
| --- | --- |
| BM25 (Query, Clarification Question + Option2) | **CosinSimilarity_TF-IDF (Query, Clarification Question + Option1)** |
| BM25 (Query, Clarification Question + Option1) | BM25 (Query, Clarification Question + Option4) |
| CosinSimilarity_TF-IDF (Query, Clarification Question + Option2) | BM25 (Query, Clarification Question + Option5) |
| **CosinSimilarity_TF-IDF (Query, Clarification Question + Option1)** | BM25 (Related Search, Clarification Question + Option5) |
| The clarification is a Statement or a Question | **BM25 (Clarification Pane, Query)** |
| **BM25 (Clarification Pane, Query)** | BM25 (Query, Clarification Question + Option3) |
| CosineSimilarity_TF-IDF (Clarification Pane, Query) | BM25 (Video Title, Clarification Question + Option5) |
| Overall Matching Terms (Clarification Pane, Query) | CosineSimilarity_TF-IDF (Video Title, Clarification Question + Option4) |
| Overall Matching Terms (Snippet, Clarification Question + Option1) | Overall Matching Terms (Related Search, Clarification Question + Option5) |
| BM25 (Video Description, Clarification Question + Option2) | BM25 (Video Title, Clarification Question + Option1) |

through Human Intelligence Tasks (HIT) on Amazon Mechanical Turk[4] (AMT) and Qualtrics[5] to gather labels.

## 4.1 Data Sampling from MIMICS-ClickExplore

The over-arching aim was to create a comprehensive dataset that can be used for generating clarification questions and re-ranking multiple clarification panes for a given query. We used the MIMICS-ClickExplore dataset that contains the corresponding aggregated user interaction signals (i.e., impression level, engagement level, and conditional click probability) for queries with multiple clarification panes. As the first selection criterion, we discarded the queries that had two clarification panes, as they are not good candidates for ranking clarification panes and are not helpful for establishing any relationship between online and offline evaluations. The query length (number of words in each query) in this dataset varied between 1 and 9. To create a new diverse search clarification dataset, we divided the queries and related clarification panes into nine subclasses based on the query length. Next, we subdivided all queries in each bin of query length based on the highest engagement level obtained by one of the associated clarification panes. After the data pre-processing, described in Subsection 3.1.1, every query had one clarification pane that had the highest engagement level compared to other panes in the set, and this highest level varied between 1 to 10 (e.g. if the highest engagement level of a clarification pane was one, then the engagement level of others for a given query was zero). Finally, we created the MIMICS-Duo dataset that contained almost 11% from each query length bin and 10% from each engagement level bin, depending on availability. Also, wherever was possible, we selected query-clarification pairs that had different impression levels. This process led to a collection of 306 queries with at least three clarification panes (1,034 query-clarification pairs) that had diversity in query length and engagement level. This dataset has the same format as the MIMICS dataset for simplicity in any analyses and comparisons in the future. The statistics of MIMICS-Duo dataset are presented in Table 6. In order to have a representative dataset, we attempted to select queries with the highest diversity in terms of engagement level, impression level, options and number of options in their clarification panes.

Table 6: Statistics of MIMICS-Duo dataset.

| | |
| --- | --- |
| Number of queries | 306 |
| Number of query-clarification pair | 1,034 |
| Number of clarification per query | 3.38±0.68 |
| Min & max clarifications per query | 3 & 8 |
| Number of candidate answer | 3.59±1.2 |
| Min & max number of candidate answers | 2 & 5 |

## 4.2 Task Design

We designed three tasks to collect judgements from AMT workers on clarification panes related to queries and search engine results pages. We then analysed the correlation between collected labels and the engagement level of clarification questions and the click-through rate of candidate answers. The tasks were designed to capture overall clarification pane preference and their quality and characteristics. Figure 2 shows an overview of the three tasks in this study.

Since the entire process was conducted online, it was necessary to prepare the instruction of each task in plain English, which was fully digestible for any worker with any level of education, and avoid academic and, in particular, information retrieval terms. We provided the required information about the survey's aim, steps that needed to be taken, and the number of questions.

AMT workers were redirected to Qualtrics to complete the tasks. This was to ensure we created a professional and user-friendly interface for the tasks. Each task had five components *(i)* the informed consent including IRB approval number and participant information sheet, *(ii)* the instruction, *(iii)* the survey body (i.e., the task itself), *(iv)* a feedback page, and *(v)* a completion code generator on the last page.

*4.2.1 Task 1 (Offline Rating): Clarification panes preferences.* We provided workers with a query and its top eight retrieved document summaries provided in MIMICS-ClickExplore. Workers were then presented with multiple (varied between 3 to 8 depending on the query) clarification panes for that query and asked to rate them using a 5-star rating. We aimed to simulate online user clicks in our task by showing all generated clarification panes for a given query at once. So the workers could rate them based on their preferences.

Finally, this included an attention check: before the workers were asked to rate the clarification panes, we showed them all clarification panes and asked them to write down the number of clarification panes that had been generated for the given query. If a worker gave an invalid answer, their HIT was rejected, and the worker was blocked from completing further HITs. In total, 306 queries with multiple clarification panes (306 HITs) were launched on AMT.

---

[4] https://www.mturk.com
[5] https://www.qualtrics.com

#### 4.2.2 Task 2 (Quality Labelling): Overall quality of clarification pane (i.e., clarification questions and their candidate answers).
Workers were shown a query and its top eight retrieved document summaries provided in the MIMICS-ClickExplore dataset. A single clarification pane was shown to workers, and they were asked to rate the *overall quality* of that clarification pane, as well as its individual candidate answers. This task was analogous to the ratings made in the MIMICS-Manual dataset, however, recall that the overlap of those queries with MIMICS-ClickExplore was insufficient to enable meaningful exploration of the relationship between user engagement and the quality of clarification panes. In our annotation process, a clarification question or a candidate answer was assessed on a 5-level rating scale (1 (very bad), 2 (bad), 3 (fair), 4 (good), and 5 (very good)). Similar to Task 1, This task also had two attention check questions to ensure a high level of quality for each HIT. In total, 1,034 query-clarification pairs (1,034 HITs) were launched on AMT.

#### 4.2.3 Task 3 (Aspect Labelling): Specific quality measures of clarification panes.
Workers were again provided with a query and its top eight retrieved document summaries, together with a single clarification pane and asked to judge the *Coverage, Diversity, Understandably*, and *Candidate Answer Order* of the clarification pane. While the MIMICS-ClickExplore dataset showed that all clarification panes generated for a given query did not receive the same user engagement level, it doesn't provide information to explore the characteristics that may lead to these differences. To address this critical question, we carried out *Aspect Labelling*. Since the *Bing* search engine generated clarification panes in a multi-choice question format and not in full sentences, we wanted to investigate whether the clarification pane was understandable and whether the presented candidate answers were diverse enough and covered all possible query intents or not. Also, we wanted to explore whether the order of the candidate answers was important or not. The overall goal was to gather data about key characteristics of clarification panes, support research into their relationship with engagement levels, and be able to generate more engaging clarification panes. The findings can also support the re-ranking clarification panes. Workers rated each aspect on a 5-level scale: strongly disagree, somewhat disagree, neither agree nor disagree, somewhat agree and strongly agree to answer the following questions, based on seeing the query and the top eight retrieved documents:

(1) Does the clarification pane have a high coverage for the given query?
(2) Does the clarification pane have a high diversity for the given query?
(3) Is the clarification pane understandable for the given query?
(4) Does the clarification pane have the correct order for the given query?

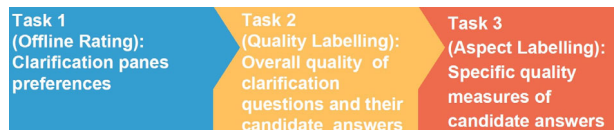

**Figure 2: An overview of the three steps of the data collection.**

To ensure that the workers understood the definition of each aspect, we presented them with a descriptive example for each question, showing clarification panes with high and low coverage, high and low diversity, understandable and non-understandable clarification panes, and with and without correct orders[6]. The feedback obtained from workers during both pilot runs and the main surveys confirmed that the task and examples were clear enough to make the justification easy for them. This task also had two gold questions, described in more detail below, to keep the quality of the collected data as high as possible. In total, 1,034 query-clarification pair (1,034 HITs) were launched on AMT.

### 4.3 Pilot Tasks
We launched two series of AMT pilot surveys, containing 9 HITs for Task *Offline Rating*, 32 HITs for Tasks *Quality Labelling*, and 32 HITs for Task *Aspect Labelling*. These pilots enabled us to analyse the flow of the tasks, estimate the required time to finish each task, collect the workers' feedback, check the quality of collected data, and revise the tasks if needed. For instance, we optimised the layout, task examples, and attention check questions (gold questions) with the aim of high validity throughout the tasks, which led to a high success rate of 89%, 91% and 100% for *Offline Rating*, *Quality Labelling* and *Aspect Labelling*, respectively, at the end of the second pilot survey.

### 4.4 Quality Assurance and Attention Measures
We embedded four quality assurance and attention measures in the task. First, to ensure workers paid attention to the different aspects of the query and the document summaries, we showed eight relevant summaries and one irrelevant summary. Workers were then asked to identify the irrelevant document summary, which was placed in a random rank position for each task. This first check provided both an attention measure (i.e., workers were forced to inspect all summaries) and a gold question (i.e., a question with a pre-defined answer). Second, we randomly inserted a second gold question from a pool of 15 pre-defined questions (e.g., *What is 2+2? Please choose five from the choices below.*). Third, we incorporated a robot detection step (CAPTCHA[7]) in each task. Lastly, workers were provided with a randomly generated code at the end of the task, which they were asked to submit to AMT as a final quality check. Answers of workers who did not pass these gold questions were removed and are not included in the final dataset. Furthermore, workers who failed the checks were also blocked from completing further tasks.

---
[6]The full instructions and examples presented to participants are available at https://github.com/Leila-Ta/MIMICS-Duo
[7]CAPTCHA (Completely Automated Public Turing test to tell Computers and Humans Apart) is a type of security measure known as challenge-response authentication.

We performed regular quality checks throughout the data collection process to ensure high-quality data, and after collecting the data, we manually checked 10% of submitted HITs per task as a final quality assurance check. If we observed any invalid submissions, we removed those submissions, prevented the workers from completing subsequent tasks, and opened the HITs to the different workers. The rigorous task design and continuous quality checks of submitted HITs helped us collect high-quality labels.

### 4.5 Crowdsourcing

This study was carried out using the AMT crowdsourcing platform between 27 January 2022 and 16 February 2022. Workers with the following qualifications were able to participate in the study:

- Only participants located in Australia, Canada, Ireland, New Zeland, the United Kingdom and the United States with a HIT approval rate of 95% or higher and a minimum of 5,000 previously approved HITs were allowed to participate in order to maximise the survey success rate and the likelihood that users were native English speakers or had a high level of English.
- Users could only participate once in each task.
- All three tasks were launched at different times and days to maximise the diversity of the participants.
- Based on experience from the pilot tasks, the hits conducted by participants who took less than 90 seconds to complete the full task were labelled as low quality and removed from the dataset, and the workers were not eligible for future tasks.

Each HIT was assigned to at least three different AMT workers. Depending on the task, the workers were paid 0.45, 0.72 and 0.95 USD per HIT for *Offline Rating*, *Quality Labelling* and *Aspect Labelling*, respectively. The collection of this dataset cost 9,880 USD. For each labelling task, we used majority voting to aggregate the annotation. In case of disagreements, the HIT was opened again to more workers until a final majority vote label could be assigned. The mean agreement was 73.44%, 74.36% and 76.63% for *Offline Rating*, *Quality Labelling* and *Aspect Labelling*.

## 5 DATA ANALYSIS

This section analyses the dataset for relationships between user engagement level and clarification pane characteristics.

### 5.1 Online Ranking vs. Offline Rating

In the task *Offline Rating*, we showed all clarification panes for a given query to the workers and asked them to rate each pane considering other panes to simulate online click behaviour as a first step in comparing online and offline evaluations. We then ranked the clarification panes for each query based on these ratings. We also ranked clarification panes based on overall quality labels collected from workers in task *Quality Labelling*. To provide a third comparison point for analysis, we also generated a random ranking. Finally, we also consider a "worst case" ranking by reversing the ideal ranked list based on the engagement level provided by MIMICS-ClickExplore (i.e., for instance, given a query, clarification panes *A*, *B* and *C* were ranked 1, 2 and 3 based on the engagement level (ideal ranked list) when we reversed the engagement levels, clarification panes *C*, *B* and *A* were ranked 1, 2 and 3). We then compared the ranked lists with our ideal ranked list, which was based on the engagement level using P@1 and MRR metrics.

No matter what evaluation metric was used (e.g., engagement level, quality label or offline rating), there were queries that had two or more clarification panes with the highest rank (tied with highest rank clarification panes). To eliminate the impact of tied clarification panes from the calculation of P@1 and MRR (as the clarification panes with the highest rank had to be selected randomly in case of tied panes), we removed those queries and related clarification panes from the dataset. Removing the tied clarification panes with the highest rank in the quality labelling and Offline Ranking collections left the dataset with 139 queries and 465 query-clarification pairs and left the dataset with 152 queries and 500 query-clarification pairs, respectively.

The results in Table 7 show that on the dataset with ties, offline rating or quality labelling cannot represent online ranking. In fact, it is evident that there is no agreement between the overall quality of clarification panes or even the offline rating of multiple clarification panes generated for given queries and the engagement level collected in MIMICS-ClickExplore based on the CTR. However, this table shows that our *Offline Rating* and *Quality Labelling* tasks had a noticeable agreement, although they have been done by different AMT workers. When we repeated the experiment on the dataset without any ties, we found out that the performance of offline rankings improved contrary to quality labelling, which means in terms of comparing online and offline re-ranking multiple clarification panes for a given query, *Offline Rating* approach seems to be a better methodology. This was expected as in the *Offline Rating* task, we showed the workers all generated clarification panes for a given query at once, and they had this opportunity to rate them based on their preferences.

**Table 7: Comparing ranking clarification panes for given queries based on offline rating, overall quality labels, random ranking and worst possible case with the ranked clarification panes based on the engagement level (ideal ranking) with and without ties.**

| Ranking Method | Tied highest rank panes | | Non-tied highest rank panes | |
|---|---|---|---|---|
| | P@1 | MRR | P@1 | MRR |
| Offline Rating | 0.382 | 0.604 | 0.382 | 0.637 |
| Quality Labelling | 0.363 | 0.599 | 0.273 | 0.576 |
| Random Ranker[1] | 0.332 | 0.576 | 0.309[2]   0.306[3] | 0.586[2]   0.581[3] |
| $\sigma$ | 0.026 | 0.015 | 0.038   0.024 | 0.041   0.025 |
| Worst Possible Case[4] | 0.0 | 0.437 | 0.0 | 0.307 |

[1] Random Ranker was repeated 1000 times and the mean values were reported.
[2] Random Ranker on the Offline Rating collection.
[3] Random Ranker on the Quality Labelling collection.
[4] There are different number of queries with different number of clarification panes (in the range of 3 to 8) in MIMICS-Duo.

### 5.2 Quality Labelling

The distribution of the quality labels for clarification panes (overall quality of clarification questions and their answers) and the quality labels of the individual candidate answers are shown in Table 8.

The number label assigned to each candidate answer is an index of its position within the clarification pane, counting from left to right, as shown in Figure 1. The results show that around 77% of clarification panes had *Good* or *Very Good* ratings. This means the majority of generated clarification panes for the given queries were relevant and satisfactory. We can also understand that although the quality of the majority of candidate answers was *Good* or *Very Good*, the mean quality rating of the candidate answers decreases from left to right across the clarification panes (i.e., with an increase in the position index of candidate answers).

To investigate the impact of the quality of candidate answers on the overall quality of clarification panes, we calculated the mean value of quality labels given to candidate answers of a clarification pane by the workers for every clarification pane. We found out that there was a strong correlation between the quality of candidate answers and the overall quality of clarification panes regardless of the number of candidate answers ($r$=0.708).

The distribution of the overall quality of clarification panes is shown for every engagement level bin (0 to 10) in Figure 3. We can see that regardless of the engagement level, almost 75% of clarification panes had *Good* or *Very Good* overall quality, and more than 96% of clarification panes had *Fair* or a higher quality label. This is a signal that a simple CTR as an indicator of user interaction with the clarification pane is not a strong metric to evaluate the performance of generating or asking clarification questions in search engines. This figure also indicates that generating high-quality clarification panes does not necessarily lead to more user engagement. Users showed that they could sometimes be reluctant to get engaged with high-quality clarifications, and they may also be engaged with poor quality ones. Therefore, it appears that click-through information can be noisy and biased and does not necessarily reflect the user's perception of information quality and therefore needs to be used carefully alongside other evaluation methods.

We also compared the quality labels of candidate answers with the click-through rate probability of candidate answers. While no correlation was found between offline answer labelling and online interaction ($\rho$=0.032), if we ranked the candidate answers based on their quality labels and click-through rate probability (ideal ranking), P@1 and MRR were calculated at 0.338 and 0.597, respectively.

**Table 8: Distribution of the quality label of clarification panes and their candidate answers.**

| | Statistics | | Labels[1](%) | | | | |
|---|---|---|---|---|---|---|---|
| Criterion | $\mu$ | $\sigma^2$ | 1 | 2 | 3 | 4 | 5 |
| Clarification Pane | 3.95 | 0.58 | 0.39 | 3.19 | 19.44 | 54.55 | 22.44 |
| Candidate Ans. #1 | 4.12 | 0.83 | 1.16 | 3.48 | 19.05 | 35.11 | 41.20 |
| Candidate Ans. #2 | 4.01 | 0.81 | 0.77 | 5.13 | 19.92 | 40.33 | 33.85 |
| Candidate Ans. #3 | 3.93 | 0.84 | 0.78 | 5.09 | 25.59 | 37.60 | 30.94 |
| Candidate Ans. #4 | 3.88 | 0.9 | 1.33 | 4.75 | 29.47 | 33.46 | 30.99 |
| Candidate Ans. #5 | 3.89 | 0.94 | 1.15 | 7.45 | 24.07 | 36.39 | 30.95 |

[1] Label meaning: 1 (Very Bad), 2 (Bad), 3 (Fair), 4 (Good), 5 (Very Good).

### 5.3 Aspect Labelling

To support the investigation of the relationship between the characteristics of clarification panes and engagement level, overall quality and offline rating of clarification panes, we carried out the *Aspect*

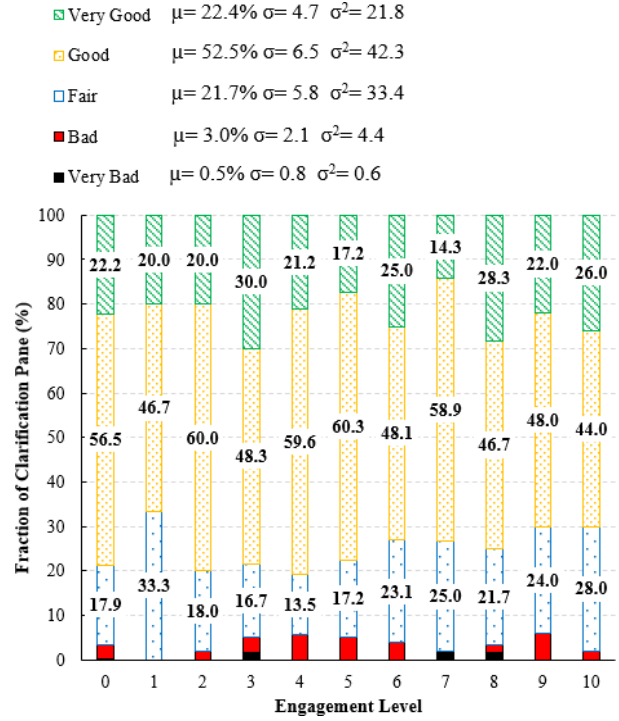

**Figure 3: Quality label vs. engagement level.**

*Labelling* task. Four aspects – *Coverage*, *Diversity*, *Understandability* and *Candidate Answer Order* – were evaluated. Table 9 shows the distribution of characteristic labels of clarification panes. It is evident that apart from the *Candidate Answer Order*, the majority of clarification panes had high *Coverage*, *Diversity* and *Understandability*, with the trend being strongest for *Understandability*. More than 40 percent of AMT workers chose the "neither agree nor disagree" response with respect to the candidate answer order aspect: here they were asked to rate whether the candidate answers for a given query were in the correct order or not (i.e. in importance order, from left to right). It appears that workers were mostly undecided regarding this aspect.

In another analysis, we classified the clarification panes into five categories based on their overall quality labels (Very Bad, Bad, Fair, Good and Very Good) and investigated the contribution of each aspect to the overall quality by calculating the mean value for each aspect in each category, shown in Figure 4. We can see the more a clarification pane had higher *Coverage* and was more *Understandable*, the higher overall quality was achieved. We can also see *Diversity* had the second place as the influential factor and *Candidate Answer Order* as mentioned, had no clear impact.

The correlations between all online and offline annotations are shown in Table 10. It is evident that the engagement level collected in MIMICS-ClickExplore (online evaluation) had no correlation with any offline measure, while different correlations can be easily found between offline measures. For example, there is a medium correlation between coverage and diversity, as expected, and overall quality or offline ranking has a higher correlation with *Coverage* compared to *Diversity* and *Understandability*. While the correlation between candidate answer order and other offline measures is very weak, the number of candidate answers also has a higher correlation with coverage and diversity compared to no correlation with understandability. This is expected as a multi-choice clarification pane can only get high coverage or diversity when the number of candidate answers is high.

Table 9: Distribution of the characteristics label of clarification panes.

| Criterion | Statistics | | Labels[1] (%) | | | | |
|---|---|---|---|---|---|---|---|
| | $\mu$ | $\sigma^2$ | 1 | 2 | 3 | 4 | 5 |
| Coverage | 3.78 | 1.18 | 3.00 | 14.02 | 12.19 | 43.23 | 27.56 |
| Diversity | 3.74 | 1.15 | 1.45 | 16.73 | 15.09 | 40.14 | 26.60 |
| Understand. | 4.61 | 0.53 | 0.39 | 2.13 | 6.09 | 18.67 | 72.73 |
| Can. Ans. Order | 3.43 | 0.87 | 1.55 | 12.86 | 40.23 | 31.62 | 13.73 |

[1] Label meaning: 1 (Strongly disagree), 2 (Somewhat disagree), 3 (Neither agree nor disagree), 4 (Somewhat agree), 5 (Strongly agree).

In another analysis shown in Table 11, we ranked clarification panes based on the *Diversity*, *Understandability* and *Candidate Answer Order* and compared the result with the ideal ranked lists based on the engagement level, quality labels and offline rating. In the first two columns, the ideal ranking is based on the online engagement level. We can see that ranking clarification panes based on the *Understandability* showed relatively higher performance compared to other aspects and even compared to ranking based on the overall quality labels (Table 7). We see the same trend when the ideal ranking is based on the offline quality labels, but when the ideal ranking is based on the offline rating, then ranking based on the *Coverage* shows the highest performance. Comparing the ranking approaches with random ranking shows that aspect ranking performed much better than random ranker when the ideal list was based on the engagement level or offline rating. However, surprisingly, random ranking outperformed all ranking approaches when the ideal ranked list was based on offline quality labels. This is another signal that offline and online evaluations on the same dataset lead to different results in search clarification.

## 6 RESEARCH ENABLED BY MIMICS-DUO

In this section, we introduce the research problems in search clarification that can be addressed using the new MIMICS-Duo dataset.
**Offline and Online Evaluation:** A key research task in search clarification is generating and asking clarification questions in information seeking problems, especially conversational search. Since MIMICS-Duo has a large query overlap with MIMICS-ClickExplore, it enables researchers and practitioners to conduct a detailed analysis of clarification selection and generation models from both online (real users) and offline (annotators) perspectives. Therefore, MIMICS-Duo complements the existing datasets for search clarification and will significantly impact the progress in this area of research.
**User Engagement and Clarification Quality:** The manual labelling of clarification panes includes information about the *coverage*, *diversity*, *understandability* of clarification panes and the *importance order* of candidate answers. This information helps the researchers to study the characteristics of clarification panes that impact user engagement.

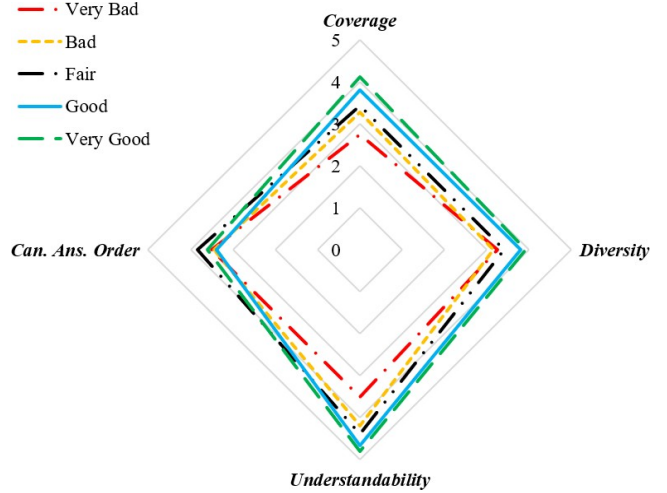

Figure 4: Mean values of different aspect labels for clarification panes with various overall quality.

**Clarification Click Models:** MIMICS-Duo contains several query-clarification pairs for a given query whose only differences are in the order of candidate answers. This information, in addition to manual annotation about the importance order of candidate answers, enables further study on training and evaluating click models for answer ranking in search clarification.

## 7 CONCLUSIONS

We introduced MIMICS-Duo, a search clarification data collection containing both online and offline evaluations. MIMICS-Duo was designed to work with the existing MIMICS-ClickExplore dataset and contains 306 unique queries with multiple clarification panes (1,034 query-clarification pairs) with interactions of real users, collected from the *Bing* search logs and graded quality labels including multiple clarification panes rating, overall quality labelling for clarification panes and their individual candidate answers and labels for different aspects of clarification panes.

Comparing online and offline evaluation is an understudied area, including in the context of search clarification. However, available search clarification datasets are either created using online user interaction signals (click-through rate) or manual annotation of quality, and there is no dataset that covers both sides. This motivated us to create the MIMICS-Duo dataset, to help bridge the gap between available search clarification datasets. This dataset was created through fine-tuned crowdsourcing, and extensive quality assurance and attention measures were considered to ensure the accuracy of the collected labels.

We analysed the relationship between the engagement level and overall quality of clarification panes and their candidate answers and investigated the characteristics of clarification panes and their impacts on the quality of clarification panes. The analysis demonstrated that the click-through rate as a signal of user engagement with clarification panes has no correlation with any offline evaluations, including the overall quality of clarification panes or offline rating. This highlights the importance of the evaluation measures used in search clarification. Although MIMICS-Duo does not compare offline evaluation with live online experimentation (e.g., real-time A/B testing with users engaging with a live system), it provides a unique opportunity for researchers to evaluate any search clarification task using offline evaluations and compare it with online signals, which was not possible before.

Table 10: Correlations between online and offline measures.

|  | Coverage | Diversity | Understandability | Can. Ans. Order | Quality Labelling | Eng. Level | Offline Rating | # of Can. Ans. |
|---|---|---|---|---|---|---|---|---|
| Coverage | NA | 0.421 | 0.313 | 0.178 | 0.227 | -0.061 | 0.273 | 0.306 |
| Diversity |  | NA | 0.260 | 0.117 | 0.176 | -0.029 | 0.245 | 0.269 |
| Understandability |  |  | NA | 0.159 | 0.226 | 0.034 | 0.227 | 0.055 |
| Can. Ans. Order |  |  |  | NA | 0.064 | 0.003 | 0.044 | -0.178 |
| Quality Labelling |  |  |  |  | NA | -0.032 | 0.225 | 0.165 |
| Eng. Level |  |  |  |  |  | NA | -0.001 | -0.079 |
| Offline Rating |  |  |  |  |  |  | NA | 0.262 |
| # of Ans. |  |  |  |  |  |  |  | NA |

Table 11: Re-ranking clarification panes using aspect labels when ideal ranking is based on engagement level, quality label or offline rating.

| Ranking Method | Eng. Level P@1 | Eng. Level MRR | Quality Labelling P@1 | Quality Labelling MRR | Offline Rating[1] P@1 | Offline Rating[1] MRR |
|---|---|---|---|---|---|---|
| Coverage | 0.333 | 0.581 | 0.317 | 0.589 | 0.369 | 0.625 |
| Diversity | 0.340 | 0.583 | 0.297 | 0.588 | 0.366 | 0.626 |
| Understandability | 0.376 | 0.601 | 0.340 | 0.608 | 0.350 | 0.613 |
| Can. Ans. Order | 0.314 | 0.565 | 0.343 | 0.604 | 0.323 | 0.584 |
| Random Ranker | 0.307 | 0.561 | 0.356 | 0.614 | 0.268 | 0.558 |

[1] Sig. dif. between coverage and random ranker, diversity and random ranker and understandability and randomranke.

In the future, we intend to propose new models for generating and asking for clarification panes that show optimum performance for both offline and online evaluations. We will also explore other features and aspects to enable us to establish a relationship between online and offline evaluations using MIMICS-Duo, including the exploration of evaluation metrics beyond P@1 and MRR for the clarification pane re-ranking task.

## ACKNOWLEDGMENTS

This research was supported in part by the Australian Research Council (DP180102687) and in part by the Center for Intelligent Information Retrieval. Any opinions, findings and conclusions or recommendations expressed in this material are those of the authors and do not necessarily reflect those of the sponsors.